\documentclass[fleqn,usenatbib]{mnras}

\usepackage{newtxtext,newtxmath}
\usepackage[T1]{fontenc}
\DeclareRobustCommand{\VAN}[3]{#2}
\let\VANthebibliography\thebibliography
\def\thebibliography{\DeclareRobustCommand{\VAN}[3]{##3}\VANthebibliography}
\usepackage{graphicx}	
\usepackage{amsmath}	
\usepackage{physics}
\usepackage{bm}
\usepackage{multirow}
\usepackage{lipsum}
\usepackage{diagbox}
\usepackage[capitalise]{cleveref}
\usepackage{tabularx}
\usepackage{subcaption}
\usepackage{hyperref}
\usepackage{float}
\usepackage{xcolor}
\title[Curvature Constraint]{Constrain Spatial Curvature and Dark Energy with Strong Lenses and Complementary Probes: a Forecast for Next-Generation Surveys}

\author[Y. Hu \& S. Dhawan]{
Yang Hu,$^{1}$\thanks{E-mail: yang.hu@astro.su.se (YH)}
Suhail Dhawan,$^{2}$\thanks{E-mail: suhail.dhawan@ast.cam.ac.uk (SD)}\\
$^{1}$Department of Astronomy, Stockholm University, Roslagstullsbacken 21, 114 21 Stockholm, Sweden\\
$^{2}$ Institute of Astronomy and Kavli Institute for Cosmology, University of Cambridge, Madingley Road, Cambridge CB3 0HA, UK
}
\begin{document}
\label{firstpage}
\pagerange{\pageref{firstpage}--\pageref{lastpage}}
\maketitle

\begin{abstract}
Inferring spatial curvature of the Universe with high-fidelity is a longstanding interest in cosmology. However, the strong degeneracy between dark energy equation-of-state parameter $w$ and curvature density parameter $\Omega_{\rm K}$ has always been a hurdle for precision measurements of curvature from late-universe probes. With the imminent commissioning of Vera C. Rubin Observatory's Legacy Survey of Space and Time (LSST), we demonstrate for the first time, using simulations of stage-IV surveys, the crucial role of time-delay distances from strong gravitational lenses in breaking this degeneracy. Our findings suggest that in non-flat $ow$CDM model, while strong lensing data alone only yield a $\Omega_{\rm K}$ constraint at $\sim\mathcal{O}(10^{-1})$ level, the integration with SNe Ia and BAO data breaks the $w$-$\Omega_{\rm K}$ degeneracy and refines the $\Omega_{\rm K}$ constraint to $\sim\mathcal{O}(10^{-2})$. This surpasses the constraints typically derived from SNe Ia Hubble diagrams and BAO data and is comparable to the constraints obtained from \textit{Planck} Primary CMB data. Additionally, we present a non-parametric approach using Gaussian Process to avoid parameter-dependency of the expansion history $H(z)$ and achieve similar $\mathcal{O}(10^{-2})$ level constraint on $\Omega_{\rm K}$. This study demonstrates the significant potential of strong gravitational lenses and Stage-IV surveys like LSST to achieve high-fidelity, independent constraints on $\Omega_{\rm K}$, contributing to our understanding of the Universe's geometry and the dynamics of dark energy. 
\end{abstract}



\section{Introduction}
The curvature density parameter $\Omega_{\rm K}$ plays a pivotal role in determining the geometric properties of the Universe, and constraining it with high-fidelity has been a longstanding interest in cosmology. A well-known endeavour to measure spatial curvature is the \textit{Planck} mission through cosmic microwave background (CMB) radiation data. The \textit{Planck} 2015 result indicated a constraint of $\abs{\Omega_{\rm K}}<0.005$ \citep{2016A&A...594A..16P}, providing evidence for a spatially flat Universe. Although constraining $\Omega_{\rm K}$ with primary CMB data along encounters limitations due to geometric degeneracy \citep[e.g.,][]{1997MNRAS.291L..33B, 10.1046/j.1365-8711.1999.02274.x, PhysRevLett.78.2054}, this degeneracy can be effectively broken by incorporating complementary probes. For instance, the \textit{Planck} 2018 result demonstrated that joint constraints on $\Omega_{\rm K}$ with baryon acoustic oscillation (BAO) measurements \citep{2020A&A...641A...6P} favored a spatially flat universe to within $\approx1\sigma$. However, recent studies have raised concerns about the consistency of such combinations, thereby raising the possibility of a ``curvature tension" in current data \citep[e.g.,][] {PhysRevD.103.L041301, Di_Valentino_2019, Di_Valentino_2021}.

The aforementioned limitations of CMB have sparked interest in using alternative, late-Universe probes that are independent of the early-Universe CMB data to constrain $\Omega_{\rm K}$. Such probes include strong gravitational lenses (Lens), type Ia supernovae (SNe~Ia) and BAO. Among these, strong gravitational lensing demands special consideration due to historical challenges in observing such phenomena, arising from their rarity and the previously limited sensitivity of astronomical instruments. First suggested by \citet{1964MNRAS.128..295R}, the Lens method involves measuring the time delays between multiple images of a distant source produced by a foreground lensing mass, from which we determine the time-delay distance $D_{\Delta t}$ of the lensing system. Earlier successful applications of Lens in cosmology include, for example, H0 Lenses in COSMOGRAIL's Wellspring \citep[H0LiCOW;][]{10.1093/mnras/stx483} and the STRong-lensing Insights into the Dark Energy Survey \citep[STRIDES;][]{10.1093/mnras/staa828}, which utilised fewer than 7 observed lensed quasars in total with measured time delays to constrain cosmological parameters. Additional contributions include studies that integrate (simulated) Lens with other late-Universe probes to constrain cosmological parameters \citep[e.g.,][] {PhysRevLett.123.231101,Qi_2022,Qi:2022sxm}. Notably, these prior studies indicate that $D_{\Delta t}$ is highly sensitive to the Hubble constant, $H_0$, yet demonstrates weak dependence on the matter density parameter $\Omega_{\rm M}$, the dark energy density parameter $\Omega_{\Lambda}$, the dark energy equation of state parameter $w$, and $\Omega_{\rm K}$ \citep[][]{Wong_2019}. 

Despite prior studies indicating a weak dependence of $D_{\Delta t}$ on $\Omega_{\rm K}$ and $w$ at face value, with the deluge of expected lensed transients expected to be discovered by the Vera C. Rubin Observatory's Legacy Survey of Space and Time (LSST), it is important to quantify what constraints are expected from LSST on parameters other than $H_0$. The survey is designed to capture multi-colour images and cover $\sim20,000$ deg$^2$ of the sky. Owing to its extensive depth and coverage, the LSST stands as the most promising survey for the observation of strongly gravitationally lensed SNe, with initial predicted numbers of several hundred discoveries annually \citep[e.g.,][]{Yoon_2019, Arendse2023, Ivezić_2019, refId0, 10.1111/j.1365-2966.2010.16639.x}. Considering these earlier predictions and LSST's 10-year survey baseline, we assume that some $1,000$ strongly lensed transients will exhibit variability detectable with LSST. This is a $\sim10^2$ order-of-magnitude improvement in event count compared to past surveys. Therefore, a method worth exploring to achieve high-fidelity constraint on $\Omega_{\rm K}$ is to combine a considerable number of $D_{\Delta t}$ data from LSST with other Stage-IV surveys involving complementary probes such as SNe~Ia and BAO.

In this paper, we use simulated Stage-IV surveys to explore the constraint on $\Omega_{\rm K}$ by 1,000 well-measured time-delay distance of Lens and complementary probes SNe~Ia and BAO. We present both a parametric and a non-parametric method to do so. In the parametric approach, an $ow$CDM model is adopted to describe the expansion history of the Universe, as this is the simplest model with non-flat geometry and also gives a degree of freedom to the equation-of-state parameter $w$ for dark energy. In the non-parametric approach, we impose a Gaussian Process (GP) prior on the expansion history so that the marginal constraint on $\Omega_{\rm K}$ is free from parametric model assumptions. Applying both of our methods to simulated Stage-IV missions data, we forecast that the data will be able to deliver $\sim \mathcal{O}(10^{-2})$ level constraints on $\Omega_\text{K}$ and a factor of $\sim2$ improvement to $w$ constraint, comparable to the current CMB-only constraints from \textit{Planck} 2018 result \citep{2020A&A...641A...6P}. 

\section{Methodology and Data}\label{sec:Methodology and data}
Strong gravitational lenses (Lens), type Ia supernovae (SNe~Ia) and baryon acoustic oscillations (BAO) are all late-Universe distance indicators. In a nutshell, Lens constrains time-delay distances, SNe~Ia constrain luminosity distances, and BAO constrains the Hubble parameter.

\subsection{Formalism}
A useful quantity in a strongly lensed system is the time-delay distances $D_{\Delta t}$ inferred from primary data on time delays and the mass distribution of lensing systems. In this study, we skip the step of inferring $D_{\Delta t}$ from primary data and directly employ simulation at the stage of time-delay distances. 

Consider a Lens system with a foreground lens at redshift $z_\text{l}$ and a distant source at redshift $z_\text{s}$ ($z_\text{s} > z_\text{l}$). The associated time-delay distance $D_{\Delta t}$ is given by:
\begin{align} \label{eq:Ddt}
    D_{\Delta t} = \frac{(1+z_\text{l})D_{\text{A,l}}D_{\text{A,s}}}{D_{\text{A,ls}}},
\end{align}

where $D_{\text{A,l}}$ and $D_{\text{A,s}}$ are angular diameter distances at lens and source respectively, and $D_{\text{A,ls}}$ is the angular diameter distance between lens and source. $D_{\text{A}}$ at a particular redshift $z$ is related to luminosity distance $D_{\text{L}}$ (which is sensitive to SNe~Ia) by $D_{\textbf{A}}=D_{\textbf{L}}/(1+z)^2$.

Complementing Lens, SNe~Ia provide measurements of the luminosity distance $D_{\text{L}}$ through the distance modulus $\mu$:
\begin{align}\label{eq:mu}
    \mu =5\lg(D_{\text{L}})+25+M,
\end{align}
where $M$ is the absolute SN~Ia calibration (also called absolute magnitude). $D_{\text{L}}$ at redshift $z$ is given by:
\begin{align} \label{eq:DL}
    D_{\text{L}} =\frac{c(1+z)}{H_0\sqrt{\abs{\Omega_{\rm K}}}}\text{sinn}\left(\sqrt{\abs{\Omega_{\rm K}}}\int_{0}^{z}\frac{\dd z'}{E(z')}\right),
\end{align}
where sinn is $\sin$ for $\Omega_{\rm K}\leq0$ and $\sinh$ for $\Omega_{\rm K}>0$. $E(z)=H(z)/H_0$ is the normalised Hubble parameter (or normalised expansion rate). $H(z)$ can be direclty measured by BAO. 

BAOs are excellent tracers of the expansion rate of the Universe as a function of redshift, and has been extensively used for inference in cosmology \citep[e.g.,][]{PhysRevD.92.123516}. Here, $H(z)$ is where parametricity of approaches comes in. In the parametric approach in this paper, we adopt an $ow$CDM model of the Universe, with $o$ allowing for non-flat geometry and $w$ being the equation-of-state parameter for dark energy\footnote{For an overview of various cosmological models, see e.g. \citet{10.1111/j.1365-2966.2012.21784.x}}. The Hubble parameter $H(z)$ takes the following form:
\begin{align}\label{eq:Hz}
    H(z)=H_0\sqrt{\Omega_{\rm M} (1+z)^3+\Omega_{\rm K}(1+z)^2+\Omega_{\text{DE}}(1+z)^{3(1+w)}},
\end{align}
where $H_0$ is the Hubble's constant, $\Omega_{\rm M}$ is the matter density parameter and $\Omega_{\text{DE}}$ is the dark energy density parameter, and $\Omega_{\text{DE}}=1-\Omega_{\rm M}-\Omega_{\rm K}$ (since we focus on late-universe probes-i.e. $z \lesssim 2$-we ignore radiation). In this case, the free cosmological parameters, labelled as $\bm{\theta}$, are $\bm{\theta}\equiv(H_0, \Omega_{\rm M}, \Omega_{\rm K}, w)$. While constraints on $w$ in a flat universe and on $\Omega_K$ assuming dark energy to be the cosmological constant (i.e. with $w=-1$) have been inferred from various probes \citep[e.g.,][]{Brout2022}, here, we focus on the $ow$CDM model. This is because with future surveys, we will be able to simultaneously constrain the parameters which would prevent a biased estimation of either one of them due to assumptions on the other.  In the non-parametric approach which we will elaborate on in \cref{sec:prior}, we do not assume \cref{eq:Hz} or any parametric models, and $H(z)$ at arbitrary $z$ is given by Gaussian Process regression instead.

\subsection{Data and Likelihood}\label{sec:likelihood}
For time-delay lenses, we use the compilation of simulated LSST redshift data of strongly lensed SNe from \citet{Goldstein_2019}. The dataset\footnote{A summary of the datasets of all three probes can be found in the \texttt{Data} folder at \url{https://github.com/YangHu99/CurvatureConstraint}} consists of 310 pairs of simulated lens and source redshifts $(z_\text{l}, z_\text{s})$. However, time-delay distances $D_{\Delta t}$ are not directly provided in this dataset. To address this, we set up a mock universe with specific cosmological parameter values for the lensed SNe: $H_0=72\ \text{km/s/Mpc}, \Omega_{\rm M}=0.3, \Omega_{\rm K}=0, w=-1$. Since we are combining the time-delay distances with the SN~Ia Hubble-Lemaitre diagram as well, we assume an SN~Ia absolute magnitude, $M$ of -19.2 mag, which is marginalised out in the inference. We label this set of mock parameters as $\bm{\theta}_0$. Using $\bm{\theta}_0$, we apply \cref{eq:Ddt,eq:DL,eq:Hz} to generate mock $D_{\Delta t}$ measurements. Additionally, we generate the corresponding percentage uncertainty $\sigma_{\text{Lens}}$ for each mock $D_{\Delta t}$ measurement, drawn randomly from a uniform distribution between 6\% to 10\%. This corresponds to the range of uncertainties in the measured time-delay distances for the existing sample of lensed transients \citep{Wong_2019} and is the expectation for future observations of lensed SNe \citep{Arendse2023}. 

We simulate 1,000 lensed transients with the expected precision. Obtaining 1000 time-delay distances with the adopted 6\%-10\% percentage errors would require substantial observational campaigns, including both spectroscopic follow-up and long-term monitoring efforts, as well as detailed modelling on time-delay distances. Nevertheless, this goal aligns with the expected capabilities of the LSST survey if we take into account all the lensed transients detectable by LSST. For lensed SNe Ia, \cite{Arendse2023} estimated a number of $\sim44$ per year or $\sim10$ `golden sample’ per year, using the LSST baseline v3.0 cadence. For lensed QSO in catalog in \cite{10.1111/j.1365-2966.2010.16639.x}, \cite{Taak_2023} estimated that $\sim1000$ would exhibit variability detectable by LSST. Taking both into consideration, achieving the required data would likely take the entire 10-year span of the LSST baseline. Techniques for detailed follow-up have been developed and applied on large sample of lensed quasars in the past \citet[e.g.,][]{Eigenbrod2005}. 

We collect the generated time-delay distance measurements and uncertainties into a data vector denoted as $\textbf{d}_{\text{Lens}}$. The log-likelihood for Lens data is then given by (up to an additive constant):
\begin{align}\label{eq:data_Lens}
    \text{ln}P(\textbf{d}_{\text{Lens}}|\bm{\theta})=-\frac{1}{2}\sum_i(\hat{D}_{\Delta t,i}-D_{\Delta t,i}(\bm{\theta}))^2/\sigma_{\text{Lens},i}^2,
\end{align}
where $\hat{D}_{\Delta t}$ is the mock time-delay distance measurement in the Universe with parameters $\bm{\theta}_0$ and $D_{\Delta t}(\bm{\theta})$ is the time-delay distance at the same redshifts but measured in another Universe with parameters $\bm{\theta}$. 

To study whether the redshift distribution in LSST data has an impact on the inference, we also adopt two alternative simulated redshift distributions as comparisons. Firstly, we generate another 1,000 pairs of $(z_{\text{l}},z_{\text{s}})$ from uniform distributions such that $z_{\text{lens}}\in[0.2,0.8]$ and $z_{\text{source}}\in[1.2z_{\text{lens}},1.4]$. We call this set of data $\text{Lens}_{\text{U}}$. Secondly, we also adopt the compilation from \citet{Arendse2023} which implemented the baseline v3.0 observing strategy. It provides 5,000 pairs of $(z_\text{l}, z_\text{s})$ and we randomly select 1,000 of them, denoting this set as $\text{Lens}_{\text{A}}$.

For complementary probe SNe~Ia, the simulated next-generation survey used in this paper is Nancy Grace Roman Space Telescope (Roman), and we use the compilation from \citet{2018ApJ...867...23H}. The dataset includes redshifts $z$ and a systematics-marginalised covariance matrix $\textbf{C}_{\text{SN}}$ for 40 binned distance moduli measurementsthe. The typical number of SNe expected in each high-z bin used in this sample from Roman is about 100 \citep[][]{2018ApJ...867...23H}. We generate mock distance modulus measurements at the 40 redshifts using \cref{eq:mu,eq:DL,eq:Hz} with mock parameters $\bm{\theta}_0$. To account for the variation in the absolute magnitude of each SN Ia event (as known as intrinsic magnitude scattering), we add a small number $\epsilon$ randomly drawn from a Gaussian distribution with $\mu=0 \ \text{mag}$ and $\sigma=0.02 \ \text{mag}$ to $M$ so that $M=-19.2+\epsilon\ \text{mag}$. Since we use binned distances for the simulated observables, given the expectation of $\sim100$ SNe per bin we take a conservative scatter of $0.2/\sqrt{100}$ mag for generating the distance moduli. This means the calibration for each SN Ia is slightly different but on average $M=-19.2$. We group the mock distance modulus measurements as a data vector $\hat{\bm{\mu}}$. Together with $\textbf{C}_{\text{SN}}$, they form the data $\textbf{d}_{\text{SN}}$, with log-likelihood given by:
\begin{align}\label{eq:data_SNe}
    \text{ln}P(\textbf{d}_{\text{SN}}|\bm{\theta},M)=-\frac{1}{2}[\hat{\bm{\mu}}-\bm{\mu}(\bm{\theta},M)]^T\textbf{C}_{\text{SN}}^{-1}[\hat{\bm{\mu}}-\bm{\mu}(\bm{\theta},M)],
\end{align}
where $\bm{\mu}(\bm{\theta},M)$ represents the distance modulus at the same redshifts in another Universe with parameter values $\bm{\theta}$ and calibration $M$.

For complementary probe BAO, we use simulated data from the Dark Energy Spectroscopic Instrument (DESI) compiled in \citet{DESI_2016}. The dataset includes redshifts in $0.05\leq z\leq1.85$ range and percentage uncertainties $\sigma_{\text{BAO}}$ in $H(z)$ for 18 binned measurements. According to table 2.4 and table 2.6 in \citet{DESI_2016} with the assumption that DESI covers $\sim9000$ deg$^2$ of the sky, the percentage uncertainties of $H(z)$ has a significant redshift dependence. A range of values between 15.09\% (for $z=0.05$) and 1.52\% (for $z=0.85$) was adopted (see also \cref{fig:Hz_interpolation}). Similar to the previous cases, we generate mock $H(z)$ measurements in the Universe with parameters $\bm{\theta}_0$ using \cref{eq:Hz}. We group the mock measurements and uncertainties into a data vector $\textbf{d}_{\text{BAO}}$. In the parametric approach, the log-likelihood for BAO data is given by:
\begin{align} \label{eq:data_BAO}
    \text{ln}P(\textbf{d}_{\text{BAO}}|\bm{\theta})=-\frac{1}{2}\sum_i(\hat{H}_i-H_i(\bm{\theta}))^2/\sigma_{\text{BAO},i}^2,
\end{align}
where $\hat{H}$ is the mock expansion rate measurement in the Universe with parameters $\bm{\theta}_0$ and $H(\bm{\theta})$ is the expansion rate at the same redshift but measured in another Universe with parameter values $\bm{\theta}$.

\subsection{Prior}\label{sec:prior} 
In Bayesian statistics, priors play a crucial role in obtaining the posterior distributions of parameters. The assumptions on priors are where the parametric approach and non-parametric approach in this paper start to diverge.

In the parametric approach, as we specifically adopt \cref{eq:Hz} as the parametric model for the expansion history $H(z)$, we need to impose priors on all cosmological parameters $H_0$, $\Omega_{\rm M}$, $\Omega_{\rm K}$, $w$ and absolute SN~Ia calibration $M$. We impose conservative uniform priors on them, and the ranges of  these priors are shown in \cref{table:prior}, with the mock parameter values of $\bm{\theta}_0$ positioned at about the midpoint of each range. 

\begin{table}
\centering
\caption {Priors on free cosmological parameters and absolute SN~Ia calibration. The column \textbf{Approach} indicates whether this parameter is involved in parametric or non-parametric approaches.}\label{table:prior}
\begin{tabular}{c c c c}
 \hline\hline
 \textbf{Parameter} & \textbf{Prior} & \textbf{Unit} & \textbf{Approach}\\
 \hline
 $H_0$ & U(0,150) & km s$^{-1}$ Mpc$^{-1}$ & parametric \& non-para\\ 
 $\Omega_{\rm M}$ & U(0,0.6) & - & only parametric\\  
 $\Omega_{\rm K}$ & U(-2,2) & - & parametric \& non-para\\
 $w$ & U(-2,0) & - & only parametric\\
 $M$ & U(-25,-15) & mag & parametric \& non-para\\
 \hline
\end{tabular}
\end{table}

While the parametric approach assumes a specific functional form for $H(z)$, this assumption may limit flexibility, as it forces us to model the expansion history based on predefined cosmological parameters. To address this, we also adopt a second non-parametric approach using Gaussian Process (GP) regression \citep[see e.g.][]{GP}. This allows $H(z)$ to be modelled as a function that is guided by the data, rather than being constrained by a parametric form. Since parameter $\Omega_{\rm M}$ and $w$ only appear in $H(z)$ and nowhere else, by making $H(z)$ non-parametric we are able to drop $\Omega_{\rm M}$ and $w$ in our model. Hence, the entire model is non-parametric in the sense that it does not depend on the expansion history of the universe.

In this approach, we treat $H(z)$ as an unknown function and apply GP regression to model it. The main assumption we make is that $H(z)$ is smooth and that its variations occur over a characteristic length scale in redshift. The `smoothness' and length scale are controlled by a set of hyperparameters $\bm \eta$ that will be marginalised in the final analysis. The GP allows us to describe a distribution over possible functions for $H(z)$, without assuming any specific mathematical form.

We define the GP prior for $H(z)$ as a normal distribution over functions:
\begin{align}\label{eq:prior}
P(H(z)|\bm{\eta}) = \mathcal{N}(H(z) | m(z), k_{\bm{\eta}}(z_i, z_j)),
\end{align}
where $m(z)$ is the mean function (based on prior expectations), $k_{\bm{\eta}}(z_i, z_j)$ is the kernel function which encodes how values of $H(z)$ at different redshifts are correlated, and $\bm \eta$ is the set of hyperparameter that defines $k_{\bm{\eta}}$.

For this analysis, we use the squared exponential (SE) kernel, which is a common choice for modelling smooth functions:
\begin{align}
k_{\text{SE}}(z_i, z_j) = a^2 \exp \left( \frac{-(z_i - z_j)^2}{2l^2} \right),
\end{align}
where the two hyperparamters ${\bm \eta}=(a, l)$ represent:
\begin{itemize}
    \item $a$ controls the amplitude of variations in $H(z)$,
    \item $l$ sets the characteristic length scale, which governs how quickly $H(z)$ can vary with redshift.
\end{itemize}

Now, every symbol in the $H(z)$ prior \cref{eq:prior} is defined. This prior will then enter the Bayesian inference framework with only 3 more parameters $H_0$, $\Omega_{\rm K}$ and $M$ that we do not consider as `expansion history' parameters in non-flat models. This method has been successfully applied to constrain $\Omega_{\rm K}$ using cosmic chronometers and the Pantheon SNe~Ia dataset \citep{Dhawan_2021}, achieving a constraint of $\Omega_{\rm K} = -0.03 \pm 0.26$. With simulated Stage-IV survey data, we expect to improve this constraint significantly.

In practice, we discretise $H(z)$ at specific redshift nodes so that it becomes a vector $\mathbf{H}$ and we can work on it using matrix. Then, the GP prior in \cref{eq:prior} becomes a multivariate Gaussian prior over the vector $\mathbf{H}$:
\begin{align} \label{eq:prior_H}
P(\mathbf{H}|\bm{\eta}) = \frac{1}{\sqrt{\abs{2\pi\mathbf{K}_{\bm{\eta}}}}} \exp \left( -\frac{1}{2} (\mathbf{H} - \mathbf{m})^{T} \mathbf{K}_{\bm{\eta}}^{-1} (\mathbf{H} - \mathbf{m}) \right),
\end{align}
where $\mathbf{K}_{\bm{\eta}}$ is the covariance matrix derived from the kernel. We impose uniform priors on the hyperparameters of the kernel and ensure that the redshift nodes are dense enough to allow accurate numerical integration of the luminosity distance for the SNe~Ia likelihood.

We perform the analysis using the \texttt{emcee} package for Markov Chain Monte Carlo (MCMC) sampling \citep{dan_foreman_mackey_2023_7574785}, and \texttt{george} for GP regression \citep{dan_foreman_mackey_2021_4541632}, which enables us to optimise the hyperparameters and jointly constrain the cosmological parameters.

\subsection{Joint Posterior}
In Bayesian framework, we need to sample on the posterior distribution which is the product of likelihood in \cref{sec:likelihood} and prior in \cref{sec:prior}. Since the treatment of BAO data differs completely between the parametric and non-parametric inference approaches, the expressions of joint posterior for the two approaches are also distinct. For notational simplicity, we define all the data used as $\textbf{d}\equiv(\textbf{d}_{\text{Lens}},\textbf{d}_{\text{SN}},\textbf{d}_{\text{BAO}})$.

In the parametric approach, the full joint posterior over $\bm{\theta}=(H_0, \Omega_{\rm M}, \Omega_{\rm K}, w)$ and $M$ is simply the product of the individual probes, since they are independent:
\begin{align}
    P(\bm{\theta},M|\textbf{d}) \propto & P(H_0)P(\Omega_{\rm M})P(\Omega_{\rm K})P(w)P(M) \\
    & P(\textbf{d}_{\text{BAO}}|\bm{\theta})P(\textbf{d}_{\text{SN}}|\bm{\theta},M)P(\textbf{d}_{\text{Lens}}|\bm{\theta}).\notag
\end{align}

In the non-parametric approach, we are jointly constraining parameters with hyperparameters, so the full joint posterior over $H_0$, $\Omega_{\rm K}$, $M$ and $\bm{\eta}$ is:
\begin{align}
    P(H_0,\Omega_{\rm K},M,\bm{\eta}|\textbf{d}) \propto &P(H_0)P(\Omega_{\rm K})P(M)P(\bm{\eta})\\ &P(\textbf{d}_{\text{BAO}}|\textbf{H})P(\textbf{d}_{\text{SN}}|H_0,\Omega_{\rm K},M,\textbf{H})\notag\\ &P(\textbf{d}_{\text{Lens}}|H_0,\Omega_{\rm K},\textbf{H})P(\textbf{H}|\bm{\eta}),\notag
\end{align}
where $P(\textbf{d}_{\text{Lens}}|H_0,\Omega_{\rm K},\textbf{H}), P(\textbf{d}_{\text{SN}}|H_0,\Omega_{\rm K},M,\textbf{H})$ and $P(\textbf{d}_{\text{BAO}}|\textbf{H})$ take similar form as \cref{eq:data_Lens}, \cref{eq:data_SNe},  \cref{eq:data_BAO} respectively, and $P(\textbf{H}|\bm{\eta})$ is \cref{eq:prior_H}.

We sample the joint posterior using MCMC sampling implemented in \texttt{emcee}, and marginalise over absolute calibration $M$ to get the joint constraint on $\Omega_{\rm K}$ and other cosmological parameters.

\section{Result and Analysis}
We show and analyse the result obtained in the parametric and non-parametric approaches using data from simulated Stage-IV surveys of Lens and complementary probes SNe~Ia and BAO.

\subsection{Parametric Model: \texorpdfstring{$ow$}-CDM}\label{sec:single_parametric}

\begin{figure*}
    \centering
    \begin{subfigure}[b]{0.48\textwidth}
        \includegraphics[width=\textwidth]{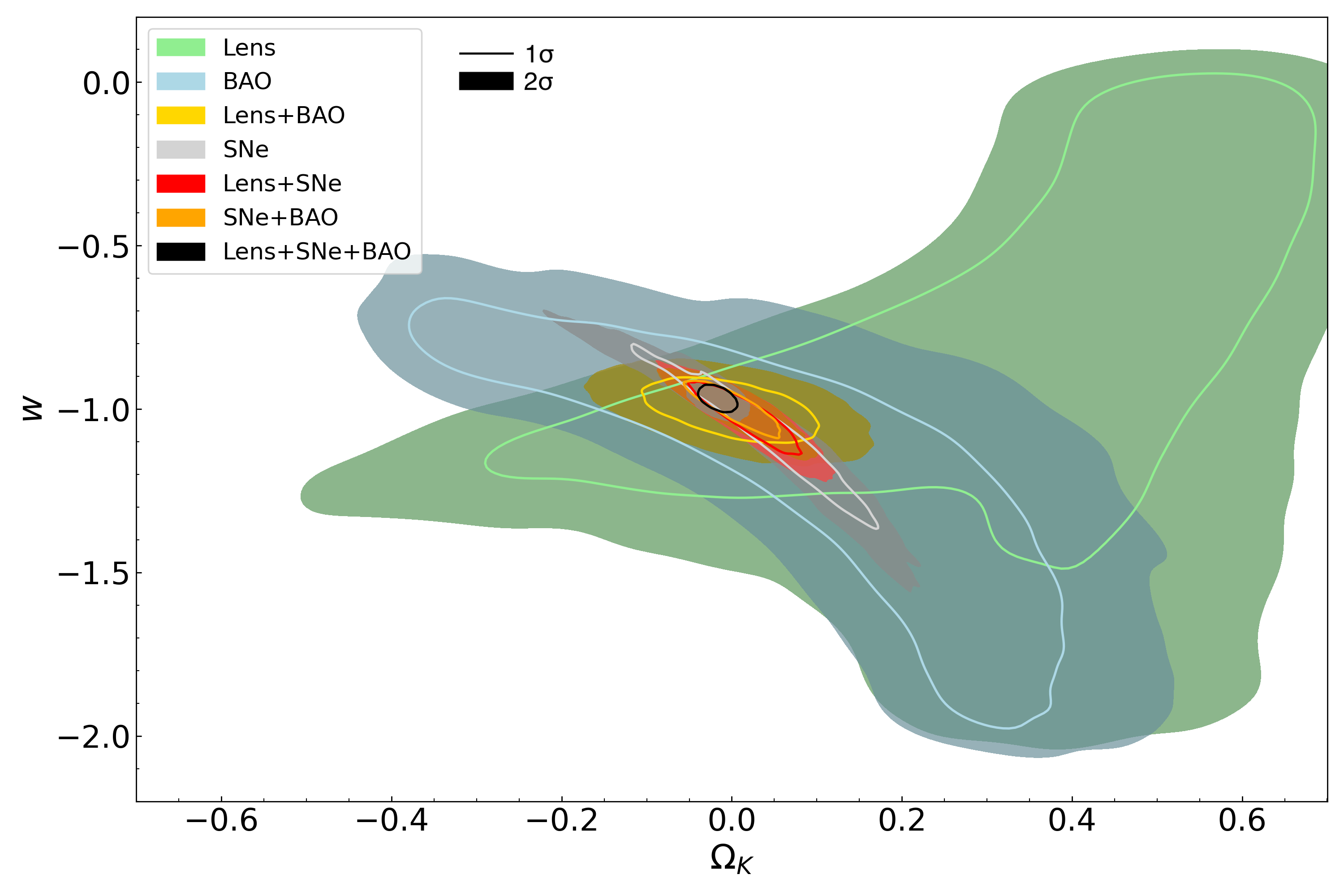}
        \caption{$w-\Omega_{\rm K}$ contour}
        \label{fig:main}
    \end{subfigure}
    \hfill
    \begin{subfigure}[b]{0.48\textwidth}
        \includegraphics[width=\textwidth]{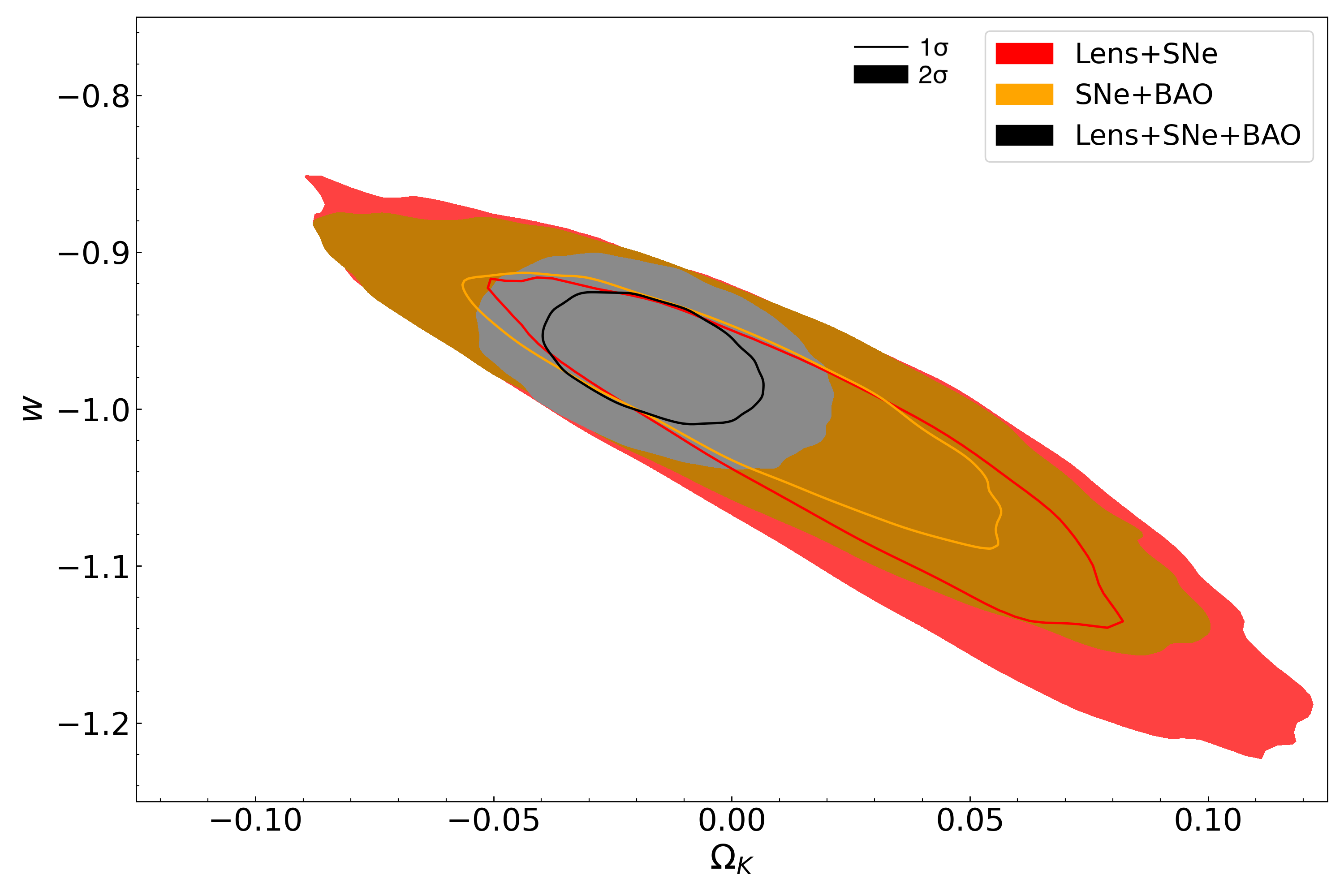}
        \caption{Zoom-in view of the three smallest contours}
        \label{fig:zoomed}
    \end{subfigure}
    \caption{$w$-$\Omega_{\rm K}$ contours for single-probe and multi-probe data. Notably, the degeneracy direction for Lens data is different from that of the SNe~Ia and BAO datasets. This breaks the degeneracy and enhancs the precision of the $2\sigma$ constraints on $\Omega_{\rm K}$ and $w$ by approximately a factor of 2.}
    \label{fig:ultimate_owCDM_contour}
\end{figure*}

We apply the parametric approach described in \cref{sec:Methodology and data} on the simulated datasets $\textbf{d}$. \cref{table:parametric} shows the result of constraints on all cosmological parameters. In \cref{table:parametric}, rows without SNe~Ia show the result of a single realisation, while rows with SNe~Ia show the arithmetic average constraints by running 3 realisations using different random seeds. This is to negate the potential bias introduced by intrinsic scattering $\epsilon$ added to $M$. Using Lens data alone, we infer $\sigma(\Omega_{\rm K})=^{+0.177}_{-0.321}$ in $ow$CDM model ($68\%$ confidence interval). This is a constraint at $\sim\mathcal{O}(10^{-1})$, similar to the non-parametric constraints with the current data \citep{Dhawan_2021}. This demonstrates that the time-delay distances by themselves are not sensitive to cosmological parameters other than $H_0$ \citep[see e.g.][]{Wong_2019}. 

We investigate the impact of combining the constraints from the lenses with the SNe~Ia and the BAO datasets from forecasts of future surveys. In \cref{table:parametric}, using Lens+SNe~Ia, we infer $\sigma(\Omega_{\rm K})=^{+0.042}_{-0.043}$ in $ow$CDM model. When combining the Lens with the SNe~Ia+BAO likelihood, we obtain $\sigma(\Omega_{\rm K}) =\pm0.015$. In comparison, the constraint from only the SNe~Ia + BAO on $\Omega_{\rm K}$ has an uncertainty of $\sim 0.04$. 

\cref{fig:ultimate_owCDM_contour} shows the degeneracy in the  $w$-$\Omega_{\rm K}$ space, which is present even when combining the SNe~Ia and BAO data.  Notably, the degeneracy direction for the Lens data differs from that observed in SNe~Ia and BAO data. The SNe~Ia and BAO datasets demonstrate a degeneracy pattern extending from the upper-left to the bottom-right, resembling a banana shape. Conversely, the Lens data exhibit a different degeneracy pattern. The $1\sigma$ contour extends from the upper-right to the middle, and then take a horizontal turn at $\sim w=-1$. This difference in degeneracy directions explains how the inclusion of the Lens data, when combined with the SNe~Ia or BAO data, effectively breaks the $w$-$\Omega_{\rm K}$ degeneracy, thereby tightening the constraint on both $w$ and $\Omega_{\rm K}$.

\begin{figure}
    \centering
    \includegraphics[width=\columnwidth]{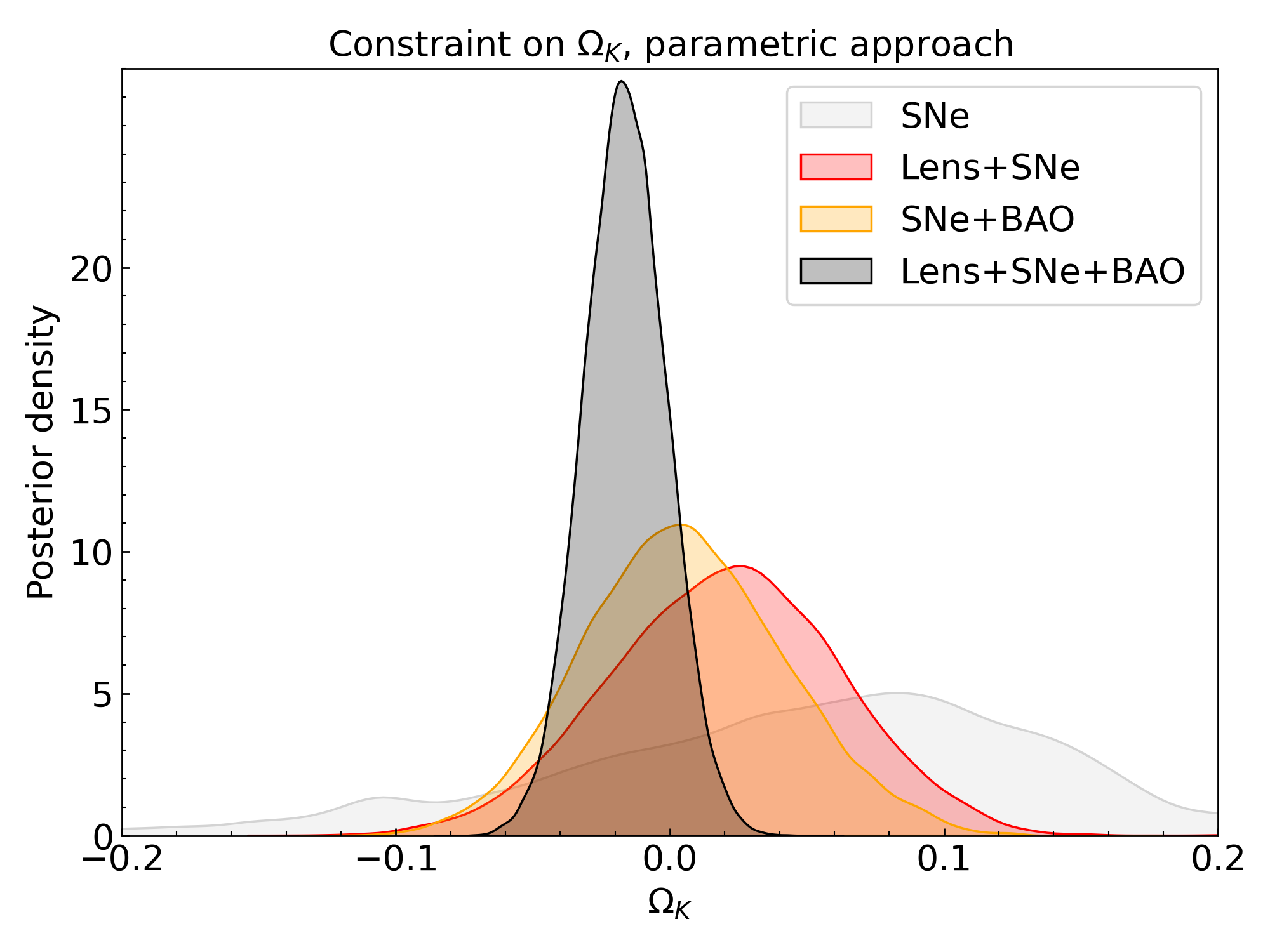}
    \caption{Kernel density estimation for $\Omega_{\rm K}$ for single-probe and combined data, assuming $ow$CDM model of the Universe, from which we infer $\sigma(\Omega_{\rm K})=^{+0.041}_{-0.043}$ for Lens+SNe~Ia and $\sigma(\Omega_{\rm K})=^{+0.038}_{-0.035}$ for SNe~Ia+BAO (68\% confidence interval).}
    \label{fig:ok_hist}
\end{figure}

The degree to which the Lens data is helping to tighten the constraint can be seen more clearly in a posterior density plot \cref{fig:ok_hist}. Here, it is evident that the Lens data significantly enhance the constraints on $\Omega_{\rm K}$ when combined with the SNe~Ia and BAO data. The $\sigma(\Omega_{\rm K})=\pm0.015$ constraint for the combined data is at $\mathcal{O}(10^{-2})$ level and is a factor of $\sim$ 2 improvement compared to the SNe~Ia Hubble diagram and BAO alone and comparable to the \textit{Planck} Primary CMB data. 

We also find that changing the source and lens redshift distribution of the Lens data does not have a significant effect on the inference, as seen from the comparison between Lens, $\text{Lens}_{\text{U}}$ and $\text{Lens}_{\text{A}}$ in the last threee rows of \cref{table:parametric}. Recall from \cref{sec:likelihood} that $\text{Lens}_{\text{U}}/\text{Lens}_{\text{A}}$ refers to the data with source and lens redshifts drawn from uniform distributions/\citet{Arendse2023} but otherwise the same as Lens.

\begin{table*}
\centering
\caption {Constraints ($68\%$ confidence interval) on various cosmological parameters in $ow$CDM model. The last three rows show comparisons between the strong lenses datasets with different redshift distributions. $\text{Lens}_{\text{U}}/\text{Lens}_{\text{A}}$ refers to the data with source and lens redshifts drawn from uniform distributions/\citet{Arendse2023} but otherwise the same as Lens.}\label{table:parametric}
\renewcommand{\arraystretch}{2} 
\begin{tabularx}{\textwidth} { 
  >{\centering\arraybackslash}X 
  >{\centering\arraybackslash}X 
  >{\centering\arraybackslash}X
  >{\centering\arraybackslash}X
  >{\centering\arraybackslash}X}
 \hline
 \hline
 \textbf{Model} & \textbf{Probe} & $\sigma(\Omega_{\rm M})$ & $\sigma(\Omega_{\rm K})$ & $\sigma(w)$\\
 \hline
 $ow$CDM & SNe~Ia& $^{+0.015}_{-0.017}$ & $^{+0.067}_{-0.082}$ & $^{+0.147}_{-0.163}$\\
 $ow$CDM & SNe~Ia+BAO & $^{+0.007}_{-0.008}$ & $^{+0.038}_{-0.036}$ & $^{+0.053}_{-0.061}$\\
 $ow$CDM & Lens & $^{+0.155}_{-0.211}$ & $^{+0.177}_{-0.321}$ & $^{+0.547}_{-0.425}$\\
 $ow$CDM & Lens+SNe~Ia & $^{+0.008}_{-0.010}$ & $^{+0.042}_{-0.043}$ & $^{+0.068}_{-0.076}$\\
 $ow$CDM & Lens+BAO & $^{+0.026}_{-0.026}$ & $^{+0.067}_{-0.067}$ & $^{+0.063}_{-0.066}$\\ 
 $ow$CDM & Lens+SNe~Ia+BAO & $^{+0.006}_{-0.006}$ & $^{+0.015}_{-0.015}$ & $^{+0.027}_{-0.028}$\\
 $ow$CDM & $\text{Lens}_{\text{U}}$+SNe~Ia+BAO & $^{+0.006}_{-0.006}$ & $^{+0.014}_{-0.014}$ & $^{+0.024}_{-0.024}$\\
 $ow$CDM & $\text{Lens}_{\text{A}}$+SNe~Ia+BAO & $^{+0.006}_{-0.006}$ & $^{+0.014}_{-0.014}$ & $^{+0.024}_{-0.025}$\\
\hline
\end{tabularx}
\end{table*}

\subsection{Non-Parametric Inference}\label{sec:Hz_non_parametric}
\begin{figure}
    \centering
    \includegraphics[width=\columnwidth]{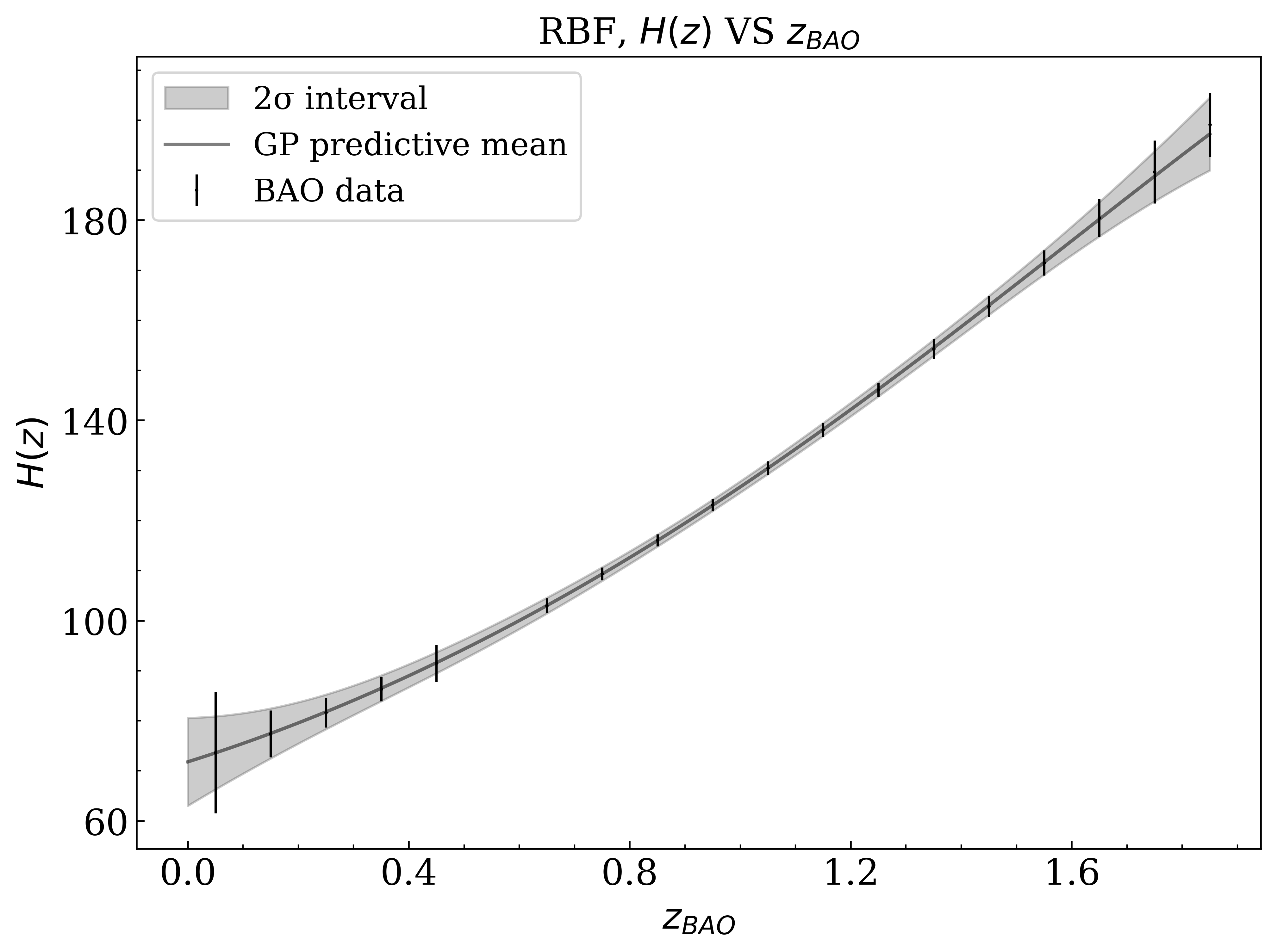}
    \caption{Non-parametric interpolation of $\textbf{H}$ with optimal hyperparameters for the RBF kernel. The solid line is the predictive mean and the grey zone indicates $2\sigma$ interval. The kernel used is RBF.}
    \label{fig:Hz_interpolation}
\end{figure}

We apply the non-parametric approach as described in \cref{sec:prior}. We first use GP with RBF kernel to interpolate the Hubble parameter $\textbf{H}$. \cref{fig:Hz_interpolation} shows an example of GP-fitted $\textbf{H}$ using a particular set of values of hyperparameter $\bm{\eta}$ in RBF. Other than RBF kernel, common choices for smooth functions include Matérn 3/2 and Matérn 5/2 kernels. Nevertheless, we found that changing the kernel to Matérn 3/2 or Matérn 5/2 does not have a significant impact on the inference.

We constrain hyperparameters to find the non-parametric inference on $\textbf{H}$, and concurrently we sample the remaining cosmological parameters by applying the same approach as the parametric case on $\textbf{d}$. Using Lens along, we infer $\sigma(\Omega_{\rm K})= \pm 0.074$. We note that while the uncertainty from the lenses is larger than the forecast for SNe \citep[see also][]{Dhawan_2021}, the lenses still provide a late-universe only constraint at the $\sim 7\%$ level. Crucially, this constraint is derived from the time-delay distance, which is a ratio of three angular diameter distances, instead of from luminosity distances to SNe~Ia.

\section{Discussion and Conclusions}
In this paper, we forecast the inference of spatial curvature with high fidelity using time-delay distance from strong gravitational lenses, complemented by distance modulus measurement from type Ia supernovae and $H(z)$ measurements from baryon acoustic oscillations. We applied the method to compiled Lens, SNe~Ia and BAO data which were respectively from simulated LSST, simulated Roman and simulated DESI, so as to forecast the constraint on $\Omega_{\rm K}$ that Stage-IV surveys will achieve. Using a parametric $ow$CDM model of the Universe, we forecast that LSST Lens data alone will reach $\sim\mathcal{O}(10^{-1})$ level constraint on $\Omega_{\rm K}$, and combining Lens data with SNe~Ia and BAO data lead to an $\mathcal{O}(10^{-2})$ level constraint on $\Omega_{\rm K}$. We also proposed a non-parametric approach to infer spatial curvature, where we GP-fitted the expansion rate $H(z)$ from simulated Stage-IV BAO data without assuming any of its parametric form. We forecast that Lens+GP-fitted $H(z)$ will also lead to $\sim\mathcal{O}(10^{-1})$ level constraint on $\Omega_{\rm K}$, and Lens+SNe~Ia+GP-fitted $H(z)$ will lead to an $\mathcal{O}(10^{-2})$ level constraint on $\Omega_{\rm K}$, consistent with the results in the parametric approach. 

For our analysis, we infer $\Omega_{\rm K}$ simultaneously with dark energy properties, e.g. the equation of state, $w$ to avoid any bias in the inference due an assumed value of $w$. Complementarity of dark energy constraints between early and late universe probes can be found in earlier studies \citep[e.g.][]{PhysRevD.84.123529} with a simplified simulation. Our work is particularly timely in light of the recent DESI 2024 results \citep[][]{DESI2024}, which underscore the significance of precisely measuring $w$, or even $w_0 w_a$, outside the standard flat $\Lambda$CDM model. Here, we demonstrate that in non-flat models, comparing to even precise measurements from a DESI-like BAO experiment, strong lenses can improve the constraints on the dark energy equation of state by a factor of $\sim$2.

We note that while constraints from time-delay lenses are themselves not very precise on $\Omega_{\rm K}$, they play a crucial role to break the $w$-$\Omega_{\rm K}$ degeneracy. In SNe~Ia and BAO data, the $w$-$\Omega_{\rm K}$ degeneracy is a key challenge that limits the precision of the constraints, even for stage-IV surveys. By including the Lens data, the $w$-$\Omega_{\rm K}$ degeneracy is effectively broken due to the difference in degeneracy directions: Lens has a unique $w$-$\Omega_{\rm K}$ contour that goes in a different direction than both SNe Ia and BAO. This enables the time-delay lenses to improve $\Omega_{\rm K}$ constraints by a factor of $\sim$2 and $w$ also by a factor of $\sim$2. As our analysis uses the number of events forecast for galaxy scale deflectors, it does not include constraints from cluster-lensed multiply imaged background sources. As demonstrated for the lensed core collapse SN~Refsdal \citep{Grillo2024}, this would further increase the precision on measuring cosmological parameters like $w$ and $\Omega_{\rm K}$, being complementary to the large sample of time-delay distances. 

The constraints on $w$ expected within the $ow$CDM model with future probes are comparable to the precision in a flat universe with combined probes using current data \citep[e.g.,][]{Brout2022}. Compared to current constraints from SNe~Ia, which are only computed with the assumption that dark energy is the cosmological constant (i.e. $w=-1$.), the constraints in the $ow$CDM model would be more than an order of magnitude more precise \citep{rubin2023union} and achieve the precision that is currently only obtained by combining late-universe probes with the CMB. While the non-parametric constraints from the strong lensing are not as precise as the expectation from SNe~Ia, however, they offer a completely independent 4$\%$ measurement of $\Omega_{\rm K}$.

With $\mathcal{O}(10^{-2})$ level constraint achieved on $\Omega_{\rm K}$ and high-fidelity constraints achieved on $w$ and other cosmological parameters, our work reveals the significant potential of late-Universe probes, particularly time-delay lenses from Stage-IV surveys, to precisely infer spatial curvature and other cosmological parameters. This, in turn, contributes to our understanding of the Universe’s geometry and the dynamics of dark energy.

\vspace{-0.2cm}
\section*{Acknowledgements}
Y. H. thanks the Institute of Astronomy, University of Cambridge for a  Summer Research Studentship when most of this research was conducted. SD acknowledges support from the Marie Curie Individual Fellowship under grant ID 890695, a Kavli Fellowship and a Junior Research Fellowship at Lucy Cavendish College. 
\section*{Data Availability}
The data and code underlying this article will be shared to the corresponding author(s) and the public. The associated repository is: 
\url{https://github.com/YangHu99/CurvatureConstraint}
\bibliographystyle{mnras}
\bibliography{reference} 

\begin{thebibliography}{}
\makeatletter
\relax
\def\mn@urlcharsother{\let\do\@makeother \do\$\do\&\do\#\do\^\do\_\do\%\do\~}
\def\mn@doi{\begingroup\mn@urlcharsother \@ifnextchar [ {\mn@doi@} {\mn@doi@[]}}
\def\mn@doi@[#1]#2{\def\@tempa{#1}\ifx\@tempa\@empty \href {http://dx.doi.org/#2} {doi:#2}\else \href {http://dx.doi.org/#2} {#1}\fi \endgroup}
\def\mn@eprint#1#2{\mn@eprint@#1:#2::\@nil}
\def\mn@eprint@arXiv#1{\href {http://arxiv.org/abs/#1} {{\tt arXiv:#1}}}
\def\mn@eprint@dblp#1{\href {http://dblp.uni-trier.de/rec/bibtex/#1.xml} {dblp:#1}}
\def\mn@eprint@#1:#2:#3:#4\@nil{\def\@tempa {#1}\def\@tempb {#2}\def\@tempc {#3}\ifx \@tempc \@empty \let \@tempc \@tempb \let \@tempb \@tempa \fi \ifx \@tempb \@empty \def\@tempb {arXiv}\fi \@ifundefined {mn@eprint@\@tempb}{\@tempb:\@tempc}{\expandafter \expandafter \csname mn@eprint@\@tempb\endcsname \expandafter{\@tempc}}}

\bibitem[\protect\citeauthoryear{{Arendse} et~al.,}{{Arendse} et~al.}{2024}]{Arendse2023}
{Arendse} N.,  et~al., 2024, \mn@doi [\mnras] {10.1093/mnras/stae1356}, \href {https://ui.adsabs.harvard.edu/abs/2024MNRAS.531.3509A} {531, 3509}

\bibitem[\protect\citeauthoryear{Aubourg et~al.,}{Aubourg et~al.}{2015}]{PhysRevD.92.123516}
Aubourg E.,  et~al., 2015, \mn@doi [Phys. Rev. D] {10.1103/PhysRevD.92.123516}, 92, 123516

\bibitem[\protect\citeauthoryear{{Bond}, {Efstathiou}  \& {Tegmark}}{{Bond} et~al.}{1997}]{1997MNRAS.291L..33B}
{Bond} J.~R.,  {Efstathiou} G.,   {Tegmark} M.,  1997, \mn@doi [\mnras] {10.1093/mnras/291.1.L33}, \href {https://ui.adsabs.harvard.edu/abs/1997MNRAS.291L..33B} {291, L33}

\bibitem[\protect\citeauthoryear{{Brout} et~al.,}{{Brout} et~al.}{2022}]{Brout2022}
{Brout} D.,  et~al., 2022, \mn@doi [\apj] {10.3847/1538-4357/ac8e04}, \href {https://ui.adsabs.harvard.edu/abs/2022ApJ...938..110B} {938, 110}

\bibitem[\protect\citeauthoryear{Collett, Montanari  \& R\"as\"anen}{Collett et~al.}{2019}]{PhysRevLett.123.231101}
Collett T.,  Montanari F.,   R\"as\"anen S.,  2019, \mn@doi [Phys. Rev. Lett.] {10.1103/PhysRevLett.123.231101}, 123, 231101

\bibitem[\protect\citeauthoryear{{DESI Collaboration} et~al.,}{{DESI Collaboration} et~al.}{2016}]{DESI_2016}
{DESI Collaboration} et~al., 2016, \mn@doi [arXiv e-prints] {10.48550/arXiv.1611.00036}, \href {https://ui.adsabs.harvard.edu/abs/2016arXiv161100036D} {p. arXiv:1611.00036}

\bibitem[\protect\citeauthoryear{{DESI Collaboration} et~al.,}{{DESI Collaboration} et~al.}{2024}]{DESI2024}
{DESI Collaboration} et~al., 2024, \mn@doi [arXiv e-prints] {10.48550/arXiv.2404.03002}, \href {https://ui.adsabs.harvard.edu/abs/2024arXiv240403002D} {p. arXiv:2404.03002}

\bibitem[\protect\citeauthoryear{Dhawan, Alsing  \& Vagnozzi}{Dhawan et~al.}{2021}]{Dhawan_2021}
Dhawan S.,  Alsing J.,   Vagnozzi S.,  2021, \mn@doi [Monthly Notices of the Royal Astronomical Society: Letters] {10.1093/mnrasl/slab058}, 506, L1

\bibitem[\protect\citeauthoryear{Efstathiou \& Bond}{Efstathiou \& Bond}{1999}]{10.1046/j.1365-8711.1999.02274.x}
Efstathiou G.,  Bond J.~R.,  1999, \mn@doi [Monthly Notices of the Royal Astronomical Society] {10.1046/j.1365-8711.1999.02274.x}, 304, 75

\bibitem[\protect\citeauthoryear{{Eigenbrod}, {Courbin}, {Vuissoz}, {Meylan}, {Saha}  \& {Dye}}{{Eigenbrod} et~al.}{2005}]{Eigenbrod2005}
{Eigenbrod} A.,  {Courbin} F.,  {Vuissoz} C.,  {Meylan} G.,  {Saha} P.,   {Dye} S.,  2005, \mn@doi [\aap] {10.1051/0004-6361:20042422}, \href {https://ui.adsabs.harvard.edu/abs/2005A&A...436...25E} {436, 25}

\bibitem[\protect\citeauthoryear{Foreman-Mackey, Bernhard, Walker, Hoyer, Kamuish, Angus  \& Mykytyn}{Foreman-Mackey et~al.}{2021}]{dan_foreman_mackey_2021_4541632}
Foreman-Mackey D.,  Bernhard J.,  Walker S.,  Hoyer S.,  Kamuish Angus R.,   Mykytyn D.,  2021, dfm/george: george v0.4.0, \mn@doi{10.5281/zenodo.4541632}, \url {https://doi.org/10.5281/zenodo.4541632}

\bibitem[\protect\citeauthoryear{Foreman-Mackey et~al.,}{Foreman-Mackey et~al.}{2023}]{dan_foreman_mackey_2023_7574785}
Foreman-Mackey D.,  et~al., 2023, dfm/emcee: emcee v3.1.4rc1, \mn@doi{10.5281/zenodo.7574785}, \url {https://doi.org/10.5281/zenodo.7574785}

\bibitem[\protect\citeauthoryear{Goldstein, Nugent  \& Goobar}{Goldstein et~al.}{2019}]{Goldstein_2019}
Goldstein D.~A.,  Nugent P.~E.,   Goobar A.,  2019, \mn@doi [The Astrophysical Journal Supplement Series] {10.3847/1538-4365/ab1fe0}, 243, 6

\bibitem[\protect\citeauthoryear{{Grillo}, {Pagano}, {Rosati}  \& {Suyu}}{{Grillo} et~al.}{2024}]{Grillo2024}
{Grillo} C.,  {Pagano} L.,  {Rosati} P.,   {Suyu} S.~H.,  2024, \mn@doi [arXiv e-prints] {10.48550/arXiv.2401.10980}, \href {https://ui.adsabs.harvard.edu/abs/2024arXiv240110980G} {p. arXiv:2401.10980}

\bibitem[\protect\citeauthoryear{Handley}{Handley}{2021}]{PhysRevD.103.L041301}
Handley W.,  2021, \mn@doi [Phys. Rev. D] {10.1103/PhysRevD.103.L041301}, 103, L041301

\bibitem[\protect\citeauthoryear{{Hounsell} et~al.,}{{Hounsell} et~al.}{2018}]{2018ApJ...867...23H}
{Hounsell} R.,  et~al., 2018, \mn@doi [\apj] {10.3847/1538-4357/aac08b}, \href {https://ui.adsabs.harvard.edu/abs/2018ApJ...867...23H} {867, 23}

\bibitem[\protect\citeauthoryear{{Huber, S.} et~al.,}{{Huber, S.} et~al.}{2019}]{refId0}
{Huber, S.} et~al., 2019, \mn@doi [A\&A] {10.1051/0004-6361/201935370}, 631, A161

\bibitem[\protect\citeauthoryear{Linder}{Linder}{2011}]{PhysRevD.84.123529}
Linder E.~V.,  2011, \mn@doi [Phys. Rev. D] {10.1103/PhysRevD.84.123529}, 84, 123529

\bibitem[\protect\citeauthoryear{Oguri \& Marshall}{Oguri \& Marshall}{2010}]{10.1111/j.1365-2966.2010.16639.x}
Oguri M.,  Marshall P.~J.,  2010, \mn@doi [Monthly Notices of the Royal Astronomical Society] {10.1111/j.1365-2966.2010.16639.x}, 405, 2579

\bibitem[\protect\citeauthoryear{{Planck Collaboration} et~al.,}{{Planck Collaboration} et~al.}{2016}]{2016A&A...594A..16P}
{Planck Collaboration} et~al., 2016, \mn@doi [\aap] {10.1051/0004-6361/201526681}, \href {https://ui.adsabs.harvard.edu/abs/2016A&A...594A..16P} {594, A16}

\bibitem[\protect\citeauthoryear{{Planck Collaboration} et~al.,}{{Planck Collaboration} et~al.}{2020}]{2020A&A...641A...6P}
{Planck Collaboration} et~al., 2020, \mn@doi [\aap] {10.1051/0004-6361/201833910}, \href {https://ui.adsabs.harvard.edu/abs/2020A&A...641A...6P} {641, A6}

\bibitem[\protect\citeauthoryear{Qi, Hu, Cui, Zhang  \& Zhang}{Qi et~al.}{2022a}]{Qi_2022}
Qi J.-Z.,  Hu W.-H.,  Cui Y.,  Zhang J.-F.,   Zhang X.,  2022a, \mn@doi [Universe] {10.3390/universe8050254}, 8, 254

\bibitem[\protect\citeauthoryear{Qi, Cui, Hu, Zhang, Cui  \& Zhang}{Qi et~al.}{2022b}]{Qi:2022sxm}
Qi J.-Z.,  Cui Y.,  Hu W.-H.,  Zhang J.-F.,  Cui J.-L.,   Zhang X.,  2022b, \mn@doi [Phys. Rev. D] {10.1103/PhysRevD.106.023520}, 106, 023520

\bibitem[\protect\citeauthoryear{{Refsdal}}{{Refsdal}}{1964}]{1964MNRAS.128..295R}
{Refsdal} S.,  1964, \mn@doi [\mnras] {10.1093/mnras/128.4.295}, \href {https://ui.adsabs.harvard.edu/abs/1964MNRAS.128..295R} {128, 295}

\bibitem[\protect\citeauthoryear{Rubin et~al.,}{Rubin et~al.}{2023}]{rubin2023union}
Rubin D.,  et~al., 2023, Union Through UNITY: Cosmology with 2,000 SNe Using a Unified Bayesian Framework (\mn@eprint {arXiv} {2311.12098})

\bibitem[\protect\citeauthoryear{Seljak \& Zaldarriaga}{Seljak \& Zaldarriaga}{1997}]{PhysRevLett.78.2054}
Seljak U. b.~u.,  Zaldarriaga M.,  1997, \mn@doi [Phys. Rev. Lett.] {10.1103/PhysRevLett.78.2054}, 78, 2054

\bibitem[\protect\citeauthoryear{Shajib et~al.,}{Shajib et~al.}{2020}]{10.1093/mnras/staa828}
Shajib A.~J.,  et~al., 2020, \mn@doi [Monthly Notices of the Royal Astronomical Society] {10.1093/mnras/staa828}, 494, 6072

\bibitem[\protect\citeauthoryear{Shi, Huang  \& Lu}{Shi et~al.}{2012}]{10.1111/j.1365-2966.2012.21784.x}
Shi K.,  Huang Y.~F.,   Lu T.,  2012, \mn@doi [Monthly Notices of the Royal Astronomical Society] {10.1111/j.1365-2966.2012.21784.x}, 426, 2452

\bibitem[\protect\citeauthoryear{Suyu et~al.,}{Suyu et~al.}{2017}]{10.1093/mnras/stx483}
Suyu S.~H.,  et~al., 2017, \mn@doi [Monthly Notices of the Royal Astronomical Society] {10.1093/mnras/stx483}, 468, 2590

\bibitem[\protect\citeauthoryear{Taak \& Treu}{Taak \& Treu}{2023}]{Taak_2023}
Taak Y.~C.,  Treu T.,  2023, \mn@doi [Monthly Notices of the Royal Astronomical Society] {10.1093/mnras/stad2201}, 524, 5446

\bibitem[\protect\citeauthoryear{Valentino, Melchiorri  \& Silk}{Valentino et~al.}{2019}]{Di_Valentino_2019}
Valentino E.~D.,  Melchiorri A.,   Silk J.,  2019, \mn@doi [Nature Astronomy] {10.1038/s41550-019-0906-9}, 4, 196

\bibitem[\protect\citeauthoryear{Valentino, Melchiorri  \& Silk}{Valentino et~al.}{2021}]{Di_Valentino_2021}
Valentino E.~D.,  Melchiorri A.,   Silk J.,  2021, \mn@doi [The Astrophysical Journal Letters] {10.3847/2041-8213/abe1c4}, 908, L9

\bibitem[\protect\citeauthoryear{Williams}{Williams}{2005}]{GP}
Williams C. E. R. . C. K.~I.,  2005, Gaussian processes for machine learning.
MIT Press

\bibitem[\protect\citeauthoryear{Wong et~al.,}{Wong et~al.}{2019}]{Wong_2019}
Wong K.~C.,  et~al., 2019, \mn@doi [Monthly Notices of the Royal Astronomical Society] {10.1093/mnras/stz3094}, 498, 1420

\bibitem[\protect\citeauthoryear{Yoon, Jee, Tyson, Schmidt, Wittman  \& Choi}{Yoon et~al.}{2019}]{Yoon_2019}
Yoon M.,  Jee M.~J.,  Tyson J.~A.,  Schmidt S.,  Wittman D.,   Choi A.,  2019, \mn@doi [The Astrophysical Journal] {10.3847/1538-4357/aaf3a9}, 870, 111

\bibitem[\protect\citeauthoryear{Željko Ivezić et~al.,}{Željko Ivezić et~al.}{2019}]{Ivezić_2019}
Željko Ivezić et~al., 2019, \mn@doi [The Astrophysical Journal] {10.3847/1538-4357/ab042c}, 873, 111

\makeatother
\end{thebibliography}
\bsp	
\label{lastpage}
\end{document}